\documentclass[runningheads]{llncs}
\usepackage[T1]{fontenc}
\usepackage{fontawesome}
\usepackage{hyperref}
\usepackage{graphicx}
\usepackage{color}

\begin{document}
\title{Multi-Domain Data Aggregation for Axon and Myelin Segmentation in Histology Images}
\titlerunning{Multi-Domain Axon and Myelin Segmentation}

 \author{Armand Collin \inst{1,2} \and
 Arthur Boschet\inst{2} \and
 Mathieu Boudreau\inst{1} \and Julien Cohen-Adad\inst{1,2}}
 \authorrunning{A. Collin, A. Boschet et al.}
 \institute{NeuroPoly Lab, Institute of Biomedical Engineering, Polytechnique Montréal, Montréal, Québec, Canada \and
 Mila - Québec Artificial Intelligence Institute, Montréal, Québec, Canada}
\maketitle              
\begin{abstract}
Quantifying axon and myelin properties (e.g., axon diameter, myelin thickness, g-ratio) in histology images can provide useful information about microstructural changes caused by neurodegenerative diseases. Automatic tissue segmentation is an important tool for these datasets, as a single stained section can contain up to thousands of axons. Advances in deep learning have made this task quick and reliable with minimal overhead, but a deep learning model trained by one research group will hardly ever be usable by other groups due to differences in their histology training data. This is partly due to subject diversity (different body parts, species, genetics, pathologies) and also to the range of modern microscopy imaging techniques resulting in a wide variability of image features (i.e., contrast, resolution). There is a pressing need to make AI accessible to neuroscience researchers to facilitate and accelerate their workflow, but publicly available models are scarce and poorly maintained. Our approach is to aggregate data from multiple imaging modalities (bright field, electron microscopy, Raman spectroscopy) and species (mouse, rat, rabbit, human), to create an open-source, durable tool for axon and myelin segmentation. Our generalist model makes it easier for researchers to process their data and can be fine-tuned for better performance on specific domains. We study the benefits of different aggregation schemes. This multi-domain segmentation model performs better than single-modality dedicated learners (p=0.03077), generalizes better on out-of-distribution data and is easier to use and maintain. Importantly, we package the segmentation tool into a well-maintained open-source software ecosystem
\footnote{\url{https://axondeepseg.readthedocs.io/}}.

\keywords{image segmentation \and histology \and axon \and myelin}
\end{abstract}
\section{Introduction}
Neurological disorders constitute the most prevalent cause of physical and cognitive disability and the second highest cause of death \cite{feigin_global_2020}. They are also a major financial burden to society, given the associated medical costs and the reduced years of employment \cite{schependom_advances_2023}. Microscopy imaging techniques play an important role to understand neurological diseases. It can notably be used to quantify demyelination and remyelination, which are critically important to assess the efficiency of new drugs.

To this end, automatic tissue segmentation is required because slices of the brain or the spinal cord, for example, can contain hundreds or thousands of axons. Typical metrics of interest include axon internal area, myelin thickness or g-ratio (ratio between inner and outer axon diameter). Collecting a meaningful amount of data cannot be done manually. As a result, researchers have been using automatic methods for more than a decade. Initial solutions consisted of a combination of thresholding \cite{more_semi-automated_2011,zhao_automatic_2010}, contour detection \cite{richerson_initial_2008,begin_automated_2014}, morphological operations \cite{zhao_automatic_2010,more_semi-automated_2011,richerson_initial_2008,zaimi_axonseg_2016}, watershed algorithms \cite{begin_automated_2014} or active contour models \cite{begin_automated_2014,zaimi_axonseg_2016}. These conventional image processing methods were effective but they relied on assumptions about the visual aspect of input images or the typical axon morphometry captured in the data \cite{more_semi-automated_2011}. These solutions required a meticulous design and were specifically tailored for a data distribution, but typically would not be applicable to other domains (i.e., different histological staining, different microscopy imaging modalities).
Deep learning approaches, more specifically convolutional neural networks (CNNs), gained a lot of popularity due to the improved performance of GPU acceleration in the last decade and large dataset sizes that have become available. These methods now outperform traditional image processing solutions for a lot of medical imaging tasks \cite{litjens_survey_2017}, including axon and myelin segmentation \cite{mesbah_deep_2016,zaimi_axondeepseg_2018}. Notably, the U-Net architecture \cite{ronneberger_u-net_2015} quickly became a \textit{de facto} standard for biomedical image segmentation, and is still widely used in the field \cite{isensee_nnu-net_2021}. For many axon and myelin segmentation methods, its encoder-decoder structure was a major design inspiration \cite{mesbah_deep_2016,janjic_measurement-oriented_2019,couedic_deep-learning_2020,deng_axondeep_2021}, and its original proposed architecture was also successfully applied to this task \cite{zaimi_axondeepseg_2018,moiseev_morphometric_2019}. Alternatively, transformers have gained a lot of traction in the deep learning community. Initially applied to language modelling, this efficient network architecture was quickly adapted for vision tasks \cite{dosovitskiy_image_2021}. An outstanding application of transformers to image segmentation is the Segment-Anything-Model \cite{kirillov_segment_2023}, a modular architecture that uses a Vision Transformer backbone. Intended to be prompted with points or bounding boxes, this model was trained on the largest annotated segmentation dataset ever released. In an effort to build a segmentation foundation model for the biomedical field, this framework was fine-tuned on various datasets (mostly CT and MRI) to create MedSAM \cite{ma_segment_2023}. Despite its promising performance on microscopy images \cite{cheng_axoncallosumem_2023,archit_segment_2023}, SAM is not ideal for axon and myelin segmentation because it heavily relies on prompts, which need to be specified for every element to segment. Since our target images often contain large quantities of axons, automating the pipeline would require to generate accurate prompts which shifts the task from segmentation to object detection.

All these conventional image processing and deep learning-based methods applied to axon and myelin segmentation share the same weakness: they were tailored for a specific image domain. As such, their performance is often impressive on the target dataset, but they perform poorly on out-of-distribution (OoD) data (i.e., different imaging modality or anatomical region). Moreover, they were often built for a specific research project, and become unmaintained a few years after the original publication. Thus, other researchers often cannot re-use existing models because these implementations are challenging to use without support from the original authors, or are not applicable to different image domains. As a result, a lot of redundant work is produced and little effort is made to make these methods easily accessible to researchers and durable in the medium- to long-term. There is a pressing need to make biomedical image segmentation models public and domain-agnostic, which is the main motivation behind this work.

 \subsection{Contribution}
We contribute a publicly available multi-domain segmentation model for axon and myelin segmentation in neurological images, trained on diverse imaging modalities, resolutions, anatomical regions, species and pathologies. We show that given a collection of datasets from multiple domains, there is no performance advantage to train dedicated models on every dataset. Aggregating the data leads to equal or improved performance on all datasets. Additionally, we demonstrate that our multi-domain model is simpler to use than single-domain methods, and its monolithic nature makes it easier to maintain. The code and weights of our open-source model can be found in a GitHub release \footnote{\url{https://github.com/axondeepseg/model_seg_generalist/releases/tag/r20240224}}. The model is also directly integrated into the AxonDeepSeg software, for a user-friendly experience with access to morphometrics extraction tools. 

\section{Methods}
\subsection{Data}
\begin{figure}
\centering
\includegraphics[width=0.75\textwidth]{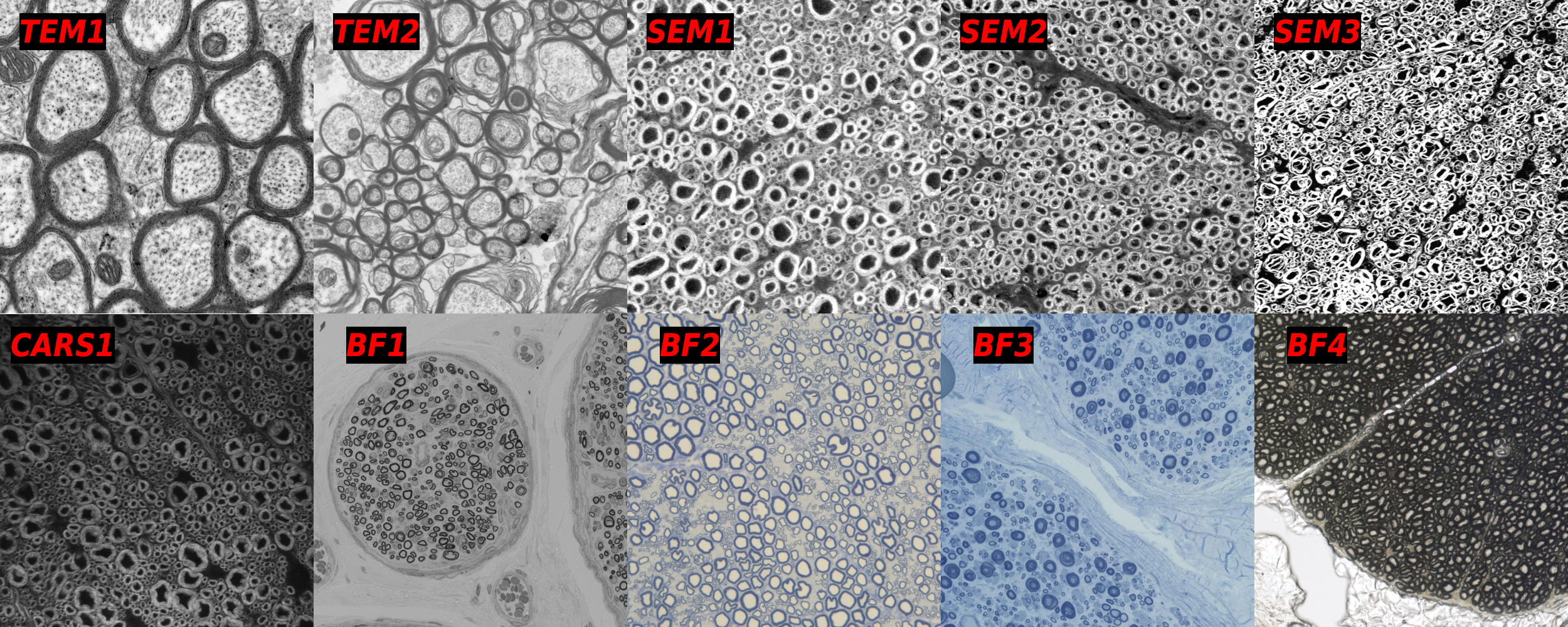}
\caption{Dataset previews} \label{fig_data_preview}
\end{figure}

\begin{table}
\centering
\caption{Dataset overview}\label{data_table}
\begin{tabular}{|l||c|c|c|c|c|c|c|c|c|c|}
\hline
\textbf{Dataset} & \texttt{TEM1} & \texttt{TEM2} & \texttt{SEM1} & \texttt{SEM2} & \texttt{SEM3} & \texttt{CARS1} & \texttt{BF1} & \texttt{BF2} & \texttt{BF3} & \texttt{BF4}\\
\hline
\textbf{modality} &  TEM & TEM & SEM & SEM & SEM & CARS & BF & BF & BF & BF\\
\hline
\textbf{annotated} & \faCheck & partially & \faCheck & partially &  & \faCheck & \faCheck & \faCheck & \faCheck &  \\
\hline
\textbf{public} & \faCheck & \faCheck & \faCheck &  & \faCheck & \faCheck & \faCheck &  &  & \faCheck \\
\hline
\textbf{species} & mouse & macaque & rat & human & dog & rat & rat & rabbit & human & cat \\
\hline
\textbf{pathology*} & H & H & H & H & H & H & H & MR & ND & H \\
 & & & & & & & MR & & & \\
\hline
\textbf{organ**} & b & b & sc & sc & sc & sc & pns & pns & b/pns/m & sc \\
\hline
\textbf{size} & 1360 & 98 & 14.8 & 31.1 & 592 & 2.6 & 280 & 12 & 20 & 658 \\
(megapixel) & & & & & & & & & & \\
\hline
\textbf{pixel size} & 0.00236 & 0.009 & 0.1 & 0.13 & 0.26 & 0.225 & 0.1 & 0.211 & 0.226 & 0.23 \\
(um/px) & & & & & & & & & & \\
\hline
\end{tabular}\\
\textbf{*} H: healthy, MR: myelin regeneration, ND: neurodegenerative diseases\\
\textbf{**} b: brain, sc: spinal cord, pns: peripheral nervous system, m: muscle
\end{table}

 \subsubsection{Datasets Used}
The datasets used in this project cover the most popular microscopy modalities: transmission electron microscopy (TEM), scanning electron microscopy (SEM), bright-field optical microscopy (BF) and the less popular coherent anti-Stokes Raman spectroscopy (CARS). Although the main focus of this work is to produce a model that performs well across modalities, the imaging technique itself only accounts for some of the variability present in the data. Subject species or pathologies change the axon morphology. The axon density is not the same in the brain, in the spinal cord or in the peripheral nervous system. Different researchers have different hardware and experimental protocols, which creates variability with all other variables controlled, depending on the provenance of the data. For example, during sample preparation, tissues are sometimes damaged or slightly deformed, which leads to artifacts in the dataset \cite{saliani_axon_2017}. All these elements come into play to affect the visual aspect of the image, and our philosophy was to include as many of these factors as possible. The aggregated dataset spans different species (rat, mouse, human, rabbit), organs (brain, spinal cord, peripheral nerves, muscles), and were acquired using four imaging modalities as previously described. A wide range of pixel sizes are present, ranging from 2.36 nm/px to 0.26 um/px, as researchers use different magnifications based on their specific needs. This diversity is summarised in Table \ref{data_table} and Figure \ref{fig_data_preview} shows visual examples. The datasets used for training were \texttt{TEM1}, \texttt{SEM1}, \texttt{CARS1}, \texttt{BF1}, \texttt{BF2} and \texttt{BF3}. Out-of-distribution evaluation was performed on datasets \texttt{TEM2}, \texttt{SEM2}, \texttt{SEM3} and \texttt{BF4}.

\subsubsection{Annotations} Regardless of the image characteristics, the task we aim to perform is shared: segmenting the axon and the myelin. As such, the ground-truth labels for this supervised 2-class segmentation task consist of axon and myelin masks. Typically, preliminary segmentations are obtained using classical image processing or deep learning based methods. The predictions are then manually corrected by annotators with various degrees of medical expertise. Occasionally, due to limited resources, it is unrealistic to collect enough masks to effectively train a model. In such cases, to alleviate the annotator’s task, an active learning strategy is employed: the model is re-trained many times, and the masks are iteratively corrected by the annotator at every step, resulting in a progressively larger training set. This strategy has the advantage of requiring less annotations, because the masks chosen for correction are targeted towards mitigating the previous model checkpoint weaknesses. Many people from different medical backgrounds were involved in this process over the last decade. We would thus expect some level of inter-rater (and even intra-rater) variability in annotation quality \cite{gros_softseg_2020,lemay_label_2023}. Although these variations are not characterised in this study because they are not deemed as problematic, our data aggregation strategy mitigates this bias. For example, a model trained on annotations with over-segmented myelin consistently reproduces this artifact in predictions, whereas a model trained on data coming from many different annotators will benefit from alternative interpretations of the data, assuming it is not overfitted.
\subsubsection{Preprocessing and Data Aggregation Strategy}
Minimal preprocessing was applied. Images were converted to grayscale when necessary, and their range was normalized to [0,1]. Every data aggregation described in this work is constructed identically. The testing set of all source datasets are combined into a large aggregated testing set. To ensure a representative validation set, we enforce the inclusion of samples from every source into the aggregated validation set. The aggregated test set is obtained by combining all source test sets.
\subsubsection{Data availability}
Most of the data used in this project came from the publicly available \textit{White Matter Microscopy Database} \cite{cohen-adad_white_2016}, namely \texttt{TEM1}, \texttt{TEM2}, \texttt{SEM1}, \texttt{SEM2}, \texttt{SEM3} and \texttt{BF4}. \texttt{CARS1} and \texttt{BF1} respectively came from \cite{duval_validation_2015} and \cite{daeschler_rapid_2022}, and are available upon request to the authors. \texttt{BF2} and \texttt{BF3} are not currently public, because the studies for which they were originally acquired are not yet published.

\subsection{Models}
 \subsubsection{Architecture and Training Details}
Two main criteria were considered to help decide the backbone for our experiments: an overall competitive performance and a durable implementation, to ensure support in the medium to long term. The latter is difficult to achieve, notably in the open-source community where project involvement and funding is often volatile. The nn-UNet framework \cite{isensee_nnu-net_2021} was selected for its consistency and popularity in the field. This project has been maintained for some years and was recently integrated into the MONAI project ecosystem \cite{cardoso_monai_2022}. As such, it seemed like the most durable option. It leverages a typical encoder-decoder U-Net architecture, which is a well-known standard for biomedical image segmentation tasks. Other alternatives were considered, including transformer-based methods \cite{kirillov_segment_2023,ma_segment_2023}, but preliminary results were not convincing and it was unclear if their implementation would still be actively maintained in the coming years. CNNs are still relevant for biomedical image segmentation because of their inherent inductive bias and they are less data-hungry than transformers \cite{dosovitskiy_image_2021}. Every model is trained based on a 5-fold cross-validation scheme for 1000 epochs. The generalist model is trained with a batch size of 13 and a patch size of 384x640. We discard the final model, which is often overfitted, and keep the checkpoint with the best validation score. All experiments were performed on a single 48 GB NVIDIA A6000 GPU.
 \subsubsection{Resolution-Ignorance}
An important design decision was to ignore the native resolution of input images. Typically, the input images fed to the network at train and test time are resampled to a common resolution, such that the model effectively works at a fixed resolution. When training on a single domain, this is not problematic because the resampling operation required to resize the train and test images is known. However, applying this model to an arbitrary image implies an appropriate resampling to the fixed internal resolution of the network. The end user needs to apply this transformation himself, or it can be done automatically based on the acquired image resolution and model target resolution. In both cases, this operation will either downsize the image, which causes information loss, or upsize it, which is computationally inefficient. Furthermore, for aggregation purposes, resampling is a liability because our data comes from a wide range of acquired resolution (spanning 2 orders of magnitude) and converting everything to the same resolution would inevitably cause catastrophic degradation in training label quality. Our proposed model is thus resolution-ignorant, as opposed to having a fixed resolution (see \cite{henschel_fastsurfervinn_2022}), but we claim its capacity is more than sufficient to efficiently generalize across scales.

\subsection{Experiments}
Two types of models are compared: dedicated learners, exclusively trained on data from a specific domain, and generalist learners, trained on aggregated data. For both experiments, we select a collection of datasets, then train a dedicated model per dataset and a generalist model on the whole collection. A visual description of our experiments is included in the appendix (see Figure \ref{fig:visual_exp_description}).

\subsubsection{Intra-Modality Aggregation}
To study the importance of intra-modality variability on model training, the intra-modality aggregation experiment uses 3 bright-field microscopy datasets (\texttt{BF1}, \texttt{BF2}, \texttt{BF3}). Despite a similar visual appearance and resolution, each dataset comes from a different species (rat, rabbit, human) and the data was acquired from multiple body parts (peripheral nervous system, brain, muscle). Additional variability comes from pathologies. Dedicated learners were trained on each dataset separately and a generalist model was trained on the concatenation of all three: \texttt{BF\_AGG}. 

\subsubsection{Inter-Modality Aggregation}
The second and most important experiment targets the impact of inter-modality variability on model performance. As such, we use datasets from 4 different modalities (\texttt{BF\_AGG}, \texttt{SEM1}, \texttt{TEM1}, \texttt{CARS1}). Note that in this context, the model trained on \texttt{BF\_AGG} is a dedicated learner, although it was considered the generalist learner of the intra-modality aggregation experiment. This task is more challenging, because the generalist model has to account for widely different image contrasts and resolutions in addition to the other factors of variability described in the previous experiment. Notably, myelin appears dark and axon light in BF/TEM images, whereas this pattern is inverted in SEM/CARS images. Moreover, the pixel sizes vary prominently, meaning that an axon with the same physical dimensions could appear to have a diameter of 10 pixels or 500 pixels depending on the magnification used. We expect the generalist model trained on the full aggregation \texttt{FULL\_AGG} to learn an even more abstract representation of the structures of interest compared to dedicated single-modality models.

\section{Results and Discussion}
We report Dice scores for all experiments in heatmaps, where every row represents a target dataset and every column a model trained on the specified source dataset. All Dice values presented are obtained by ensembling the 5 folds of the cross-validation scheme. Results for both axon and myelin classes are presented. In 3.2, the generalist model is applied to unseen data.
\subsection{Intra- and Inter-Modality Aggregation Results}
\begin{figure}
\includegraphics[width=\textwidth]{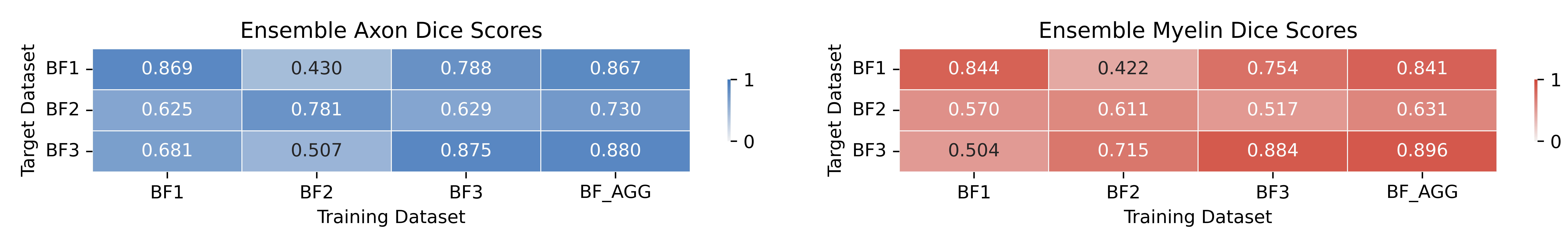}
\caption{Intra-Modality Aggregation Results: Performance of dedicated and generalist models on all BF datasets} \label{fig_intra}
\end{figure}

As shown in Figure \ref{fig_intra}, the model trained on \texttt{BF\_AGG} performs similarly to dedicated BF models. Dedicated learners generally work well across BF datasets, because these intra-modality image domains share a similar visual appearance. However, the generalist model consistently outperforms dedicated models on datasets they were not trained on.

\begin{figure}
\includegraphics[width=\textwidth]{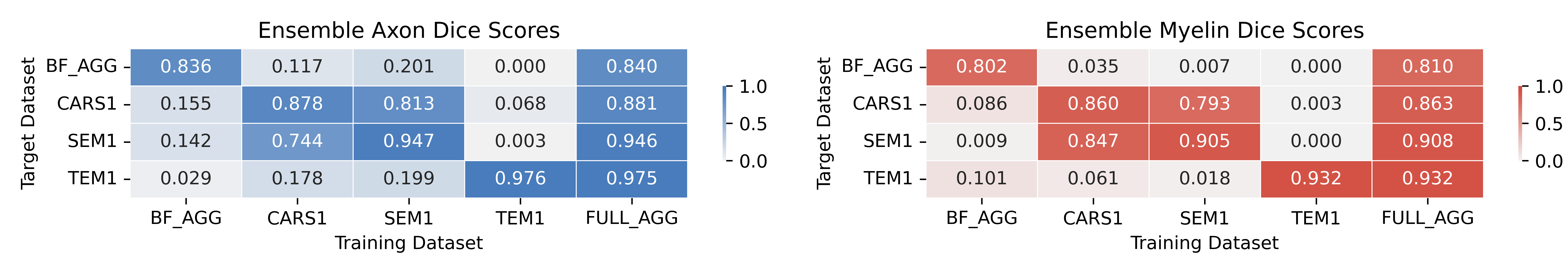}
\caption{Inter-Modality Aggregation Results: Performance of dedicated and generalist models on all imaging modalities.} \label{fig_inter}
\end{figure}

Expectedly, the heatmaps presented in Figure \ref{fig_inter} are sparse: dedicated models work poorly on image modalities they were not trained on. The only exception is the similar behavior of dedicated models trained on \texttt{CARS1} and \texttt{SEM1}, which makes sense given these two modalities are visually similar. The performance of the generalist and dedicated models for both classes are compared using a paired Student's t-test on pairs of Dice score. For a fair comparison, we only include the performance of dedicated models on datasets they were trained on. The Dice scores of the generalist model are significantly greater than the ones of dedicated models (p=0.03077, N=8).

\subsection{Out-of-Distribution Generalization}
\begin{table}[]
\centering
\caption{Dice Scores on out-of-distribution data.}\label{tab_ood}
\begin{tabular}{l|cc|cc|}
\cline{2-5}
                                 & \multicolumn{2}{c|}{SEM2}                            & \multicolumn{2}{c|}{TEM2}                            \\ \cline{2-5} 
                                 & \multicolumn{1}{c|}{Axon}           & Myelin         & \multicolumn{1}{c|}{Axon}           & Myelin         \\ \hline
\multicolumn{1}{|l|}{Dedicated}  & \multicolumn{1}{c|}{0.824}          & 0.774          & \multicolumn{1}{c|}{0.640}          & 0.604          \\ \hline
\multicolumn{1}{|l|}{Generalist} & \multicolumn{1}{c|}{\textbf{0.834}} & \textbf{0.783} & \multicolumn{1}{c|}{\textbf{0.697}} & \textbf{0.706} \\ \hline
\end{tabular}
\end{table}

Table \ref{tab_ood} compares dedicated models to the generalist model on OoD data. For \texttt{SEM2} and \texttt{TEM2}, we respectively used the SEM and TEM dedicated models. The generalist model trained on the full aggregation \texttt{FULL\_AGG} outperforms both dedicated models on these datasets. Notably, the generalist model consistently detects more small axons, possibly due to its multi-resolution training set.
Our proposed model  was also tested on unlabelled datasets \texttt{SEM3} and \texttt{BF4}. Examples of OoD predictions are included in the appendix for qualitative evaluation (see Figures \ref{fig:id_examples} and \ref{fig:ood_examples}).

\section{Conclusion}
Our proposed generalist model produces better segmentations than single modality learners on in-distribution and out-of-distribution images. Our work shows that although intra-modality aggregation is useful, inter-modality data aggregation is the most beneficial. Our strategy is more sustainable than maintaining multiple dedicated systems, and leads to a single easy-to-use model. Models trained on aggregations \texttt{BF\_AGG} and \texttt{FULL\_AGG} are publicly available. We hope this project facilitates both the workflow of neuroscience researchers and the medium- to long-term maintenance of the method.

\begin{credits}
 \subsubsection{\ackname} We would like to thank Tanguy Duval and Daniel Côté for the CARS images, Simeon Christian Daeschler, Marie-Hélène Bourget, Tessa Gordon and Gregory Howard Borschel for the \texttt{BF1} dataset, Charles R. Reiter and Geetanjanli Bendale for the \texttt{BF2} dataset, and Osvaldo Delbono for the \texttt{BF3} dataset. 

\subsubsection{\discintname}
The authors have no competing interests to declare that are relevant to the content of this article.
\end{credits}

\bibliographystyle{splncs04}
\bibliography{refs}

\appendix
\section*{Appendix}

\renewcommand{\thefigure}{A}
\begin{figure}
    \centering
    \includegraphics[width=\textwidth]{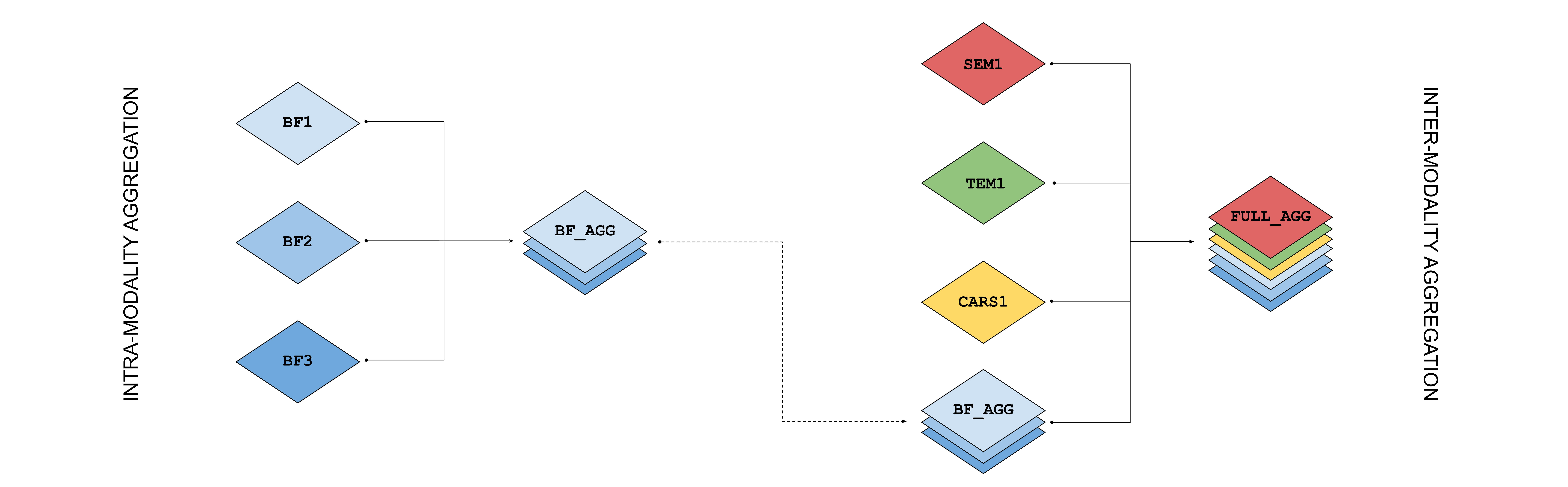}
    \caption{Visual description of experiments}
    \label{fig:visual_exp_description}
\end{figure}

\renewcommand{\thefigure}{B}
\begin{figure}
    \centering
    \includegraphics[width=\textwidth]{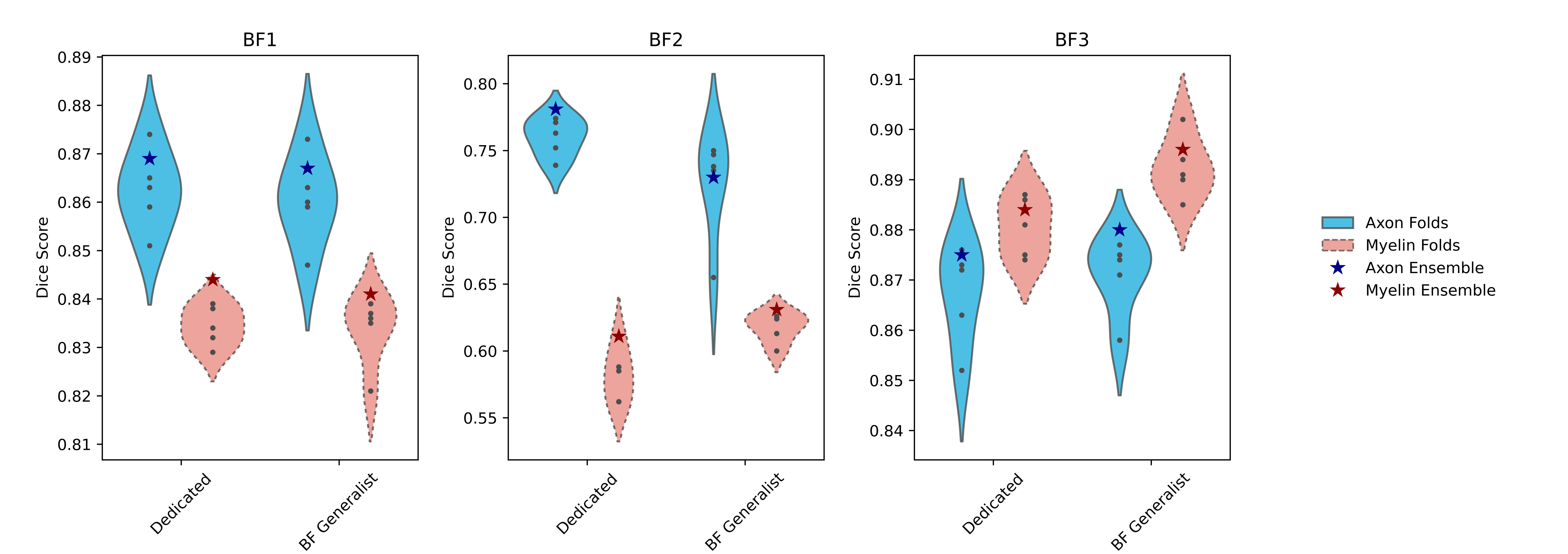}
    \caption{Dataset-wise results of individual folds for \textbf{intra-modality}.}
    \label{fig:detailed_intra}
\end{figure}

\renewcommand{\thefigure}{C}
\begin{figure}
    \centering
    \includegraphics[width=\textwidth]{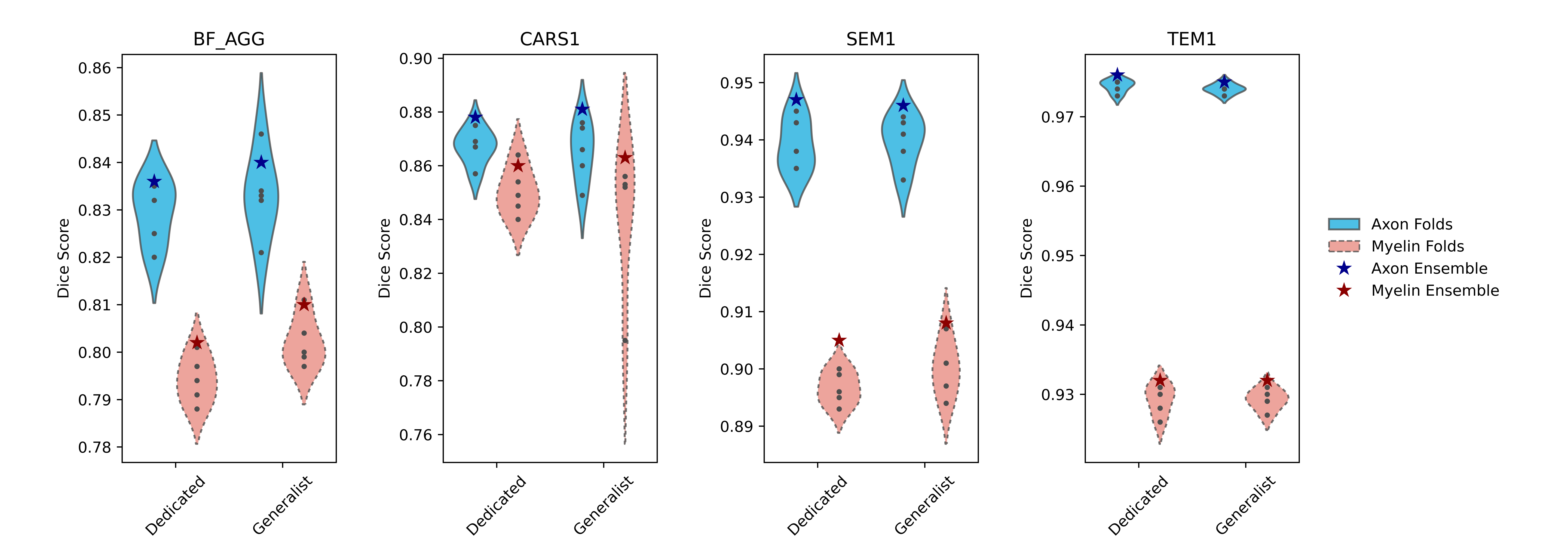}
    \caption{Dataset-wise results of individual folds for \textbf{inter-modality}.}
    \label{fig:detailed_inter}
\end{figure}

\renewcommand{\thefigure}{D}
\begin{figure}
    \centering
    \includegraphics[width=\textwidth]{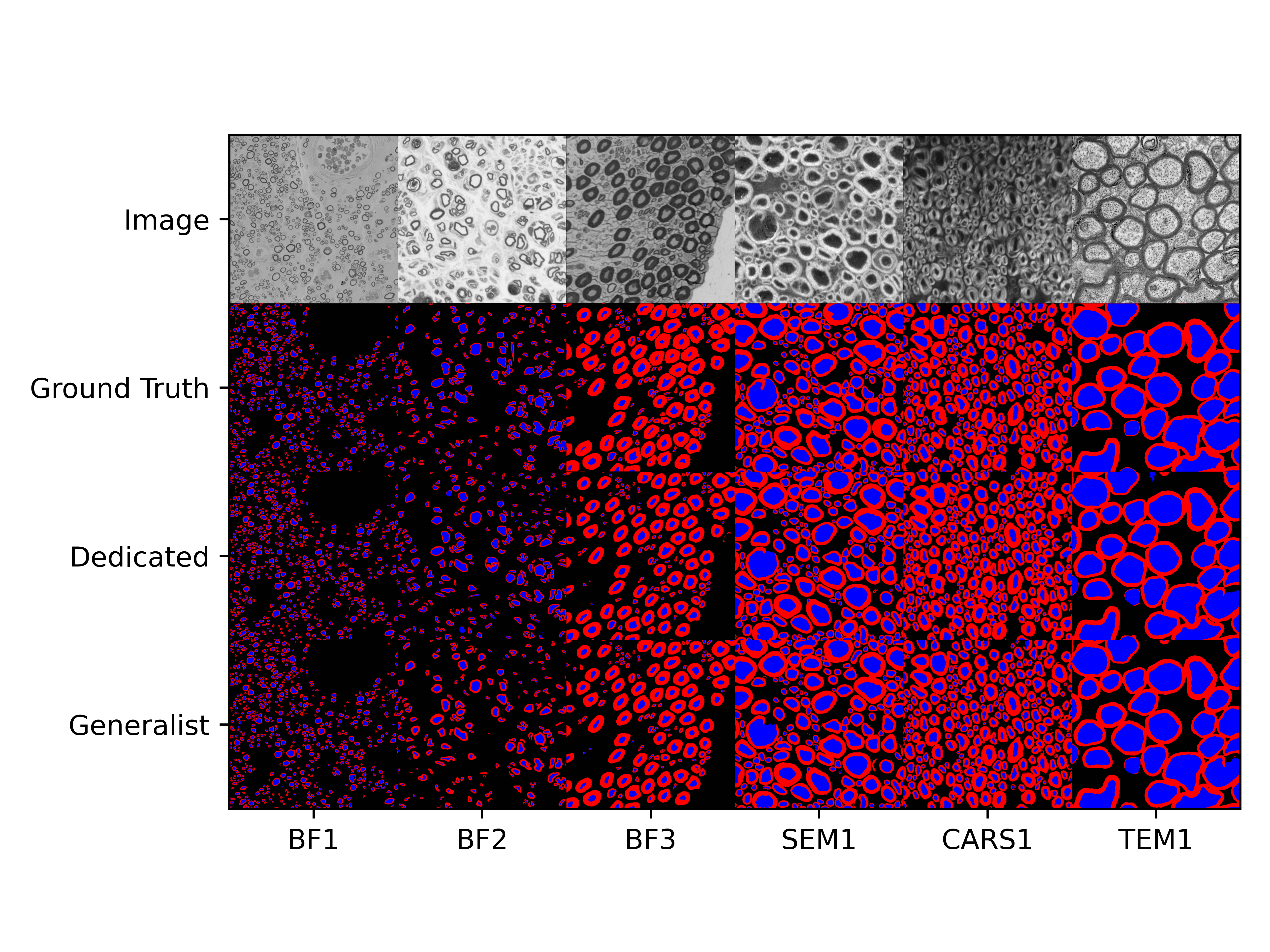}
    \caption{In-Distribution predictions. All dedicated models used for the third row were trained on the corresponding dataset specified for every column.}
    \label{fig:id_examples}
\end{figure}

\renewcommand{\thefigure}{E}
\begin{figure}
    \centering
    \includegraphics[width=\textwidth]{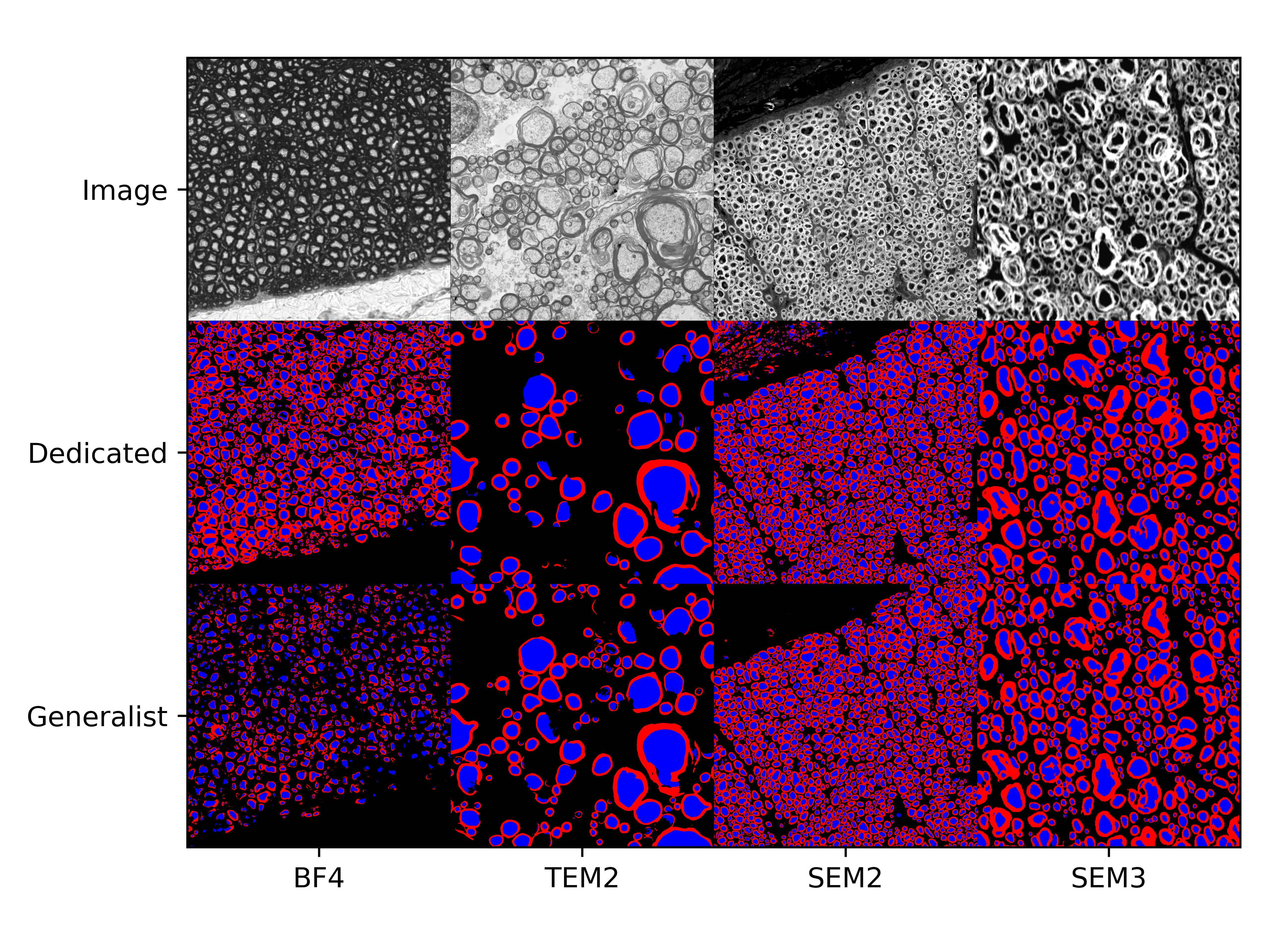}
    \caption{Out-of-Distribution predictions. The dedicated models used for the second row were respectively trained on \texttt{BF\_AGG}, \texttt{TEM1}, \texttt{SEM1} and \texttt{SEM1}. Note the remarkable performance of the BF generalist model on its OoD input.}
    \label{fig:ood_examples}
\end{figure}

\end{document}